\documentclass[iop]{emulateapj}
\usepackage{apjfonts,ulem}
\usepackage[backref,breaklinks,colorlinks,citecolor=blue,linkcolor=magenta,urlcolor=blue]{hyperref}
\usepackage{color,soul}

\begin{document}

\title{Parallel-propagating Fluctuations at Proton-kinetic Scales in the Solar Wind are Dominated by Kinetic Instabilities}%Collisionless Solar Wind Turbulence Does Not Contain Parallel-propagating Fluctuations at Proton-kinetic Scales

\author{Lloyd D.\ Woodham}
\email{woodhamlloyd@gmail.com}
\affiliation{Mullard Space Science Laboratory, University College London, Holmbury St. Mary, Surrey RH5 6NT, UK}
\affiliation{Department of Physics, The Blackett Laboratory, Imperial College London, London, SW7 2AZ, UK}

\author{Robert T.\ Wicks}
\affiliation{Mullard Space Science Laboratory, University College London, Holmbury St. Mary, Surrey RH5 6NT, UK}
\affiliation{Institute of Risk and Disaster Reduction, University College London, London WC1E 6BT, UK}

\author{Daniel Verscharen}
\affiliation{Mullard Space Science Laboratory, University College London, Holmbury St. Mary, Surrey RH5 6NT, UK}
\affiliation{Space Science Center, University of New Hampshire, Durham, NH 03824, USA}

\author{Christopher J.\ Owen}
\affiliation{Mullard Space Science Laboratory, University College London, Holmbury St. Mary, Surrey RH5 6NT, UK}

\author{Bennett A.\ Maruca}
\affiliation{Department of Physics and Astronomy, University of Delaware, Newark, DE 19716, USA}
\affiliation{Bartol Research Institute, University of Delaware, Newark, DE 19716, USA}

\author{Benjamin L.\ Alterman}
\affiliation{Department of Climate and Space Sciences and Engineering, University of Michigan, Ann Arbor, MI 48109, USA}
\affiliation{Department of Applied Physics, University of Michigan, Ann Arbor, MI 48109, USA}

\begin{abstract}
	
We use magnetic helicity to characterise solar wind fluctuations at proton-kinetic scales from \textit{Wind} observations. For the first time, we separate the contributions to helicity from fluctuations propagating at angles quasi-parallel and oblique to the local mean magnetic field, $\mathbf{B}_0$. We find that the helicity of quasi-parallel fluctuations is consistent with Alfv\'en-ion cyclotron and fast magnetosonic-whistler modes driven by proton temperature anisotropy instabilities and the presence of a relative drift between $\alpha$-particles and protons. We also find that the helicity of oblique fluctuations has little dependence on proton temperature anisotropy and is consistent with fluctuations from the anisotropic turbulent cascade. Our results show that parallel-propagating fluctuations at proton-kinetic scales in the solar wind are dominated by proton temperature anisotropy instabilities and not the turbulent cascade. We also provide evidence that the behaviour of fluctuations at these scales is independent of the origin and macroscopic properties of the solar wind.

\end{abstract}

%\maketitle

\section{\label{sec:level1}Introduction}

The solar wind is a plasma that emanates from the solar corona and expands supersonically to form the heliosphere. This dynamic environment supports fluctuations such as turbulence, waves, and instabilities over a broad range of scales \citep{Verscharen2019}. The coupling of electromagnetic fluctuations and particles over many scales is integral to energy transport and heating in plasmas. \textit{In situ} measurements of the solar wind provide insights into these fundamental processes, making it a unique plasma laboratory to better understand other astrophysical plasmas that are inaccessible to spacecraft.

Solar wind fluctuations are predominately Alfv\'enic and exhibit a turbulent cascade of energy from large to small scales that is mediated by non-linear interactions \citep{Bruno2013,Chen2016a}. At wave-numbers $k\ll 2\pi/d_p$ and $k\ll 2\pi/\rho_p$, where $d_p$ is the proton inertial length and $\rho_p$ is the proton gyroradius, the plasma behaves as a fluid. This range of scales is denoted the inertial range of turbulence and is characterised by fluctuations with increasing anisotropy ($k_\perp \gg k_\parallel$) towards smaller scales with respect to $\mathbf{B}_0$, the local mean magnetic field \citep{Wicks2010a,Chen2011,Chen2012a,Horbury2012,Lacombe2017}. At proton-kinetic scales, i.e., $k\sim 2\pi/d_p$ and $k\sim 2\pi/\rho_p$, Hall and Larmor-radius effects become important in mediating the physics of the cascade \citep{Alexandrova2013}, and the Alfv\'enic fluctuations show properties consistent with dispersive kinetic Alfv\'en waves (KAWs) \citep{Leamon1999,Bale2005,Howes2008b,Sahraoui2010}. At these scales, the turbulent fluctuations are prone to collisionless damping via wave-particle interactions, which leads to fine structure in particle velocity distribution functions (VDFs) \citep{Chen2018}. This fine structure increases the effective collision rate, enabling dissipation of the fluctuations and leading to plasma heating.

Solar wind particle VDFs often deviate from isotropic Maxwellian distributions due to a low rate of collisional relaxation \citep{Kasper2008,Marsch2012,Maruca2013a,Kasper2017}. Non-Maxwellian features such as temperature anisotropies relative to $\mathbf{B}_0$, beams, and relative drifts between plasma species provide sources of free energy for instabilities %
\citep{Kasper2002a,Hellinger2006,Kasper2008,Bale2009,Maruca2012,Bourouaine2013,Kasper2013,Gary2015a,Alterman2018}. One example is the proton temperature anisotropy, $T_{p,\perp}/T_{p,\parallel}$, where $T_{p,\perp}$ and $T_{p,\parallel}$ are the proton temperatures perpendicular and parallel to $\mathbf{B}_0$, respectively. As the solar wind flows out into the heliosphere, local processes drive changes in $T_{p,\perp}/T_{p,\parallel}$, leading to a deviation from Chew-Goldberger-Low theory for adiabatic expansion \citep{Chew1956,Matteini2007}. If $T_{p,\perp}/T_{p,\parallel}$ deviates far enough from unity, kinetic instabilities grow that act to limit this anisotropy. Measurements of the near-Earth solar wind show that the observed range of $T_{p,\perp}/T_{p\parallel}$ values is constrained by the increasing growth rates of these anisotropy-driven instabilities \citep{Kasper2002a,Hellinger2006,Bale2009,Maruca2012}. In fact, \citet{Klein2018} show that over half of solar wind intervals support ion-scale kinetic instabilities, suggesting that they are ubiquitous in the solar wind.

Four kinetic instabilities driven by proton temperature anisotropy are relevant in the solar wind. The Alfv\'en ion-cyclotron (AIC) and mirror-mode instabilities are unstable at $T_{p,\perp}$ sufficiently greater than $T_{p,\parallel}$. On the other hand, the parallel and oblique firehose instabilities are unstable at $T_{p,\parallel}$ sufficiently greater than $T_{p,\perp}$. The AIC and parallel firehose instabilities have maximum growth rates for wave-vectors, $\mathbf{k}$, that are parallel to $\mathbf{B}_0$, which respectively leads to growing AIC and fast magnetosonic-whistler (FMW) modes at $k_\parallel d_p\lesssim1$. Conversely, the mirror-mode and oblique firehose instabilities, have maximum growth rates for $\mathbf{k}$ at angles oblique to $\mathbf{B}_0$, and drive modes at $k_\perp\rho_p\lesssim1$ that do not propagate in the plasma frame. The two parallel instabilities can also be driven unstable by particle beams and drifts \citep{Bourouaine2013,Verscharen2013}, for example, the differential flow between $\alpha$-particles and protons, $\mathbf{v}_d=\mathbf{v}_\alpha-\mathbf{v}_p$ \citep{Neugebauer1994,Neugebauer1996,Steinberg1996}. This drift velocity is about $v_d\simeq 0.6\,v_A$, where $v_A$ the local Alfv\'en speed, and directed along $\mathbf{B}_0$ away from the Sun \citep{Kasper2006,Alterman2018}. \citet{Podesta2011a,Podesta2011} show that the presence of a differential flow leads to a preferential driving of the AIC and parallel firehose instabilities in the direction of $\mathbf{v}_d$ and $-\mathbf{v}_d$, respectively.

Several studies \citep{He2011,He2012a,He2012,Podesta2011,Klein2014,Bruno2015,Telloni2015} use magnetic helicity to characterise solar wind fluctuations at proton-kinetic scales. However, Taylor's hypothesis \citep{Taylor1938} limits single-spacecraft observations to the spacecraft frame, so that we can only measure a projection of $\mathbf{k}$ along the flow direction past the spacecraft, $k_r=\mathbf{k}\cdot\mathbf{v}_{sw}$, where $\mathbf{v}_{sw}$ is the solar wind velocity. In this letter, we use a novel method to measure the wave-vector anisotropy of solar wind magnetic field fluctuations using magnetic helicity \citep{Wicks2012}. For the first time, we separate the helicity of fluctuations propagating at quasi-parallel and oblique angles to $\mathbf{B}_0$. We find that periods of strong coherent helicity correspond to parallel-propagating fluctuations during intervals in which the plasma is unstable due to its proton temperature anisotropy. These fluctuations are preferentially driven due to the presence of a significant drift between $\alpha$-particles and protons. Furthermore, we show that the continual background helicity in the solar wind corresponds to fluctuations propagating oblique to $\mathbf{B}_0$. The amplitude of this signature shows little dependence on $\beta_{p,\parallel}$ and $T_{p,\perp}/T_{p,\parallel}$, and we attribute these fluctuations to the anisotropic turbulent cascade \citep{Horbury2008,Chen2010a,Wicks2010a}. Our results suggest there is no strong parallel component of the turbulent cascade at proton-kinetic scales.% \hl{and provide evidence for the universality of proton-kinetic physics in the solar wind.}

\section{\label{sec:level1A}Magnetic Helicity}

Magnetic helicity is a measure of the phase coherence between magnetic field components and serves as a useful indicator of the polarisation properties of solar wind fluctuations. The fluctuating magnetic helicity density in spectral form is defined as $H_m(\mathbf{k})\equiv\mathbf{A}(\mathbf{k})\cdot\mathbf{B}^*(\mathbf{k})$, where $\mathbf{A}$ is the fluctuating magnetic vector potential, $\mathbf{B}$ is the fluctuating magnetic field, and the asterisk indicates the complex conjugate of the Fourier coefficients \citep{Matthaeus1982b}. From a single-spacecraft time series of magnetic field measurements, we can only determine a reduced form of the magnetic helicity density \citep{Batchelor1970,Montgomery1981,Matthaeus1982a}:

\begin{equation}
H^r_m(k_r)=\frac{2\,\mathrm{Im}\left\lbrace\mathsf{P}_{TN}(k_r)\right\rbrace}{k_r},
\end{equation} %If RTN coordinates - a minus sign is needed in front of this equation

\noindent where $\mathsf{P}_{ij}(k_r)=B_i^*(k_r)\cdot B_j(k_r)$ is the reduced power spectral tensor in RTN coordinates. We define the normalised reduced fluctuating magnetic helicity density as:

\begin{equation}
\sigma_m(k_r)\equiv\frac{k_rH^r_m(k_r)}{\left|\mathbf{B}(k_r)\right|^2}=\frac{2\,\mathrm{Im}\left\lbrace\mathsf{P}_{TN}(k_r)\right\rbrace}{\mathrm{Tr}\left\lbrace\mathsf{P}(k_r)\right\rbrace},
\end{equation}

\noindent where $\mathrm{Tr}\lbrace\rbrace$ denotes the trace. Here, $\sigma_m(k_r)$ is dimensionless and takes values between $[-1,1]$, where $\sigma_m=-1$ indicates purely left-handed and $\sigma_m=+1$ indicates purely right-handed circular fluctuations, respectively. A value of $\sigma_m=0$ indicates no overall coherence. We define the field-aligned coordinate system ($\hat{x}$,$\hat{y}$,$\hat{z}$),

\begin{eqnarray}
\hat{z}=\frac{\mathbf{B}_0}{\left|\mathbf{B}_0\right|};
\:\hat{y}=-\frac{\mathbf{v}_{sw}\times\mathbf{B}_0}{\left|\mathbf{v}_{sw}\times\mathbf{B}_0\right|};
\:\hat{x}=\hat{y}\times\hat{z},
\label{eqncoords}
\end{eqnarray}

\noindent so that $\mathbf{v}_{sw}$ lies in the $\hat{x}$-$\hat{z}$ plane \citep{Wicks2012}. This coordinate system exploits Taylor's hypothesis so that we can separate the different contributions to magnetic helicity from fluctuations propagating quasi-parallel and oblique to $\mathbf{B}_0$ using the definition:

\begin{equation}
\sigma_{ij}(k_r)=\frac{2\,\mathrm{Im}\left\lbrace\mathsf{P}_{ij}(k_r)\right\rbrace}{\mathrm{Tr}\left\lbrace\mathsf{P}(k_r)\right\rbrace},
\label{eqnhelFA}
\end{equation}

\noindent where the indices $i,j=x,y,z$. Therefore, $\sigma_{xy}$ gives the helicity of fluctuations with $\mathbf{k}\times\mathbf{B}_0\simeq0$ and $\sigma_{yz}$ the helicity for fluctuations with $\mathbf{k}\times\mathbf{B}_0\neq0$. The component $\sigma_{xz}$ integrates to zero if the distribution of fluctuation power is gyrotropic. This novel analysis technique allows us to recover additional information about the wave-vector of the fluctuations using magnetic helicity, without assuming any particular linear or non-linear wave mode.

\section{Method} \label{sec:Method}

We analyse magnetic field and ion moment data from the MFI fluxgate magnetometer \citep{Lepping1995,Koval2013} and SWE Faraday cup \citep{Ogilvie1995,Kasper2006} instruments on-board the \textit{Wind} spacecraft \citep{Acuna1995} from Jun 2004 to Oct 2018. We neglect collisionally old wind, $A_c\geq1$, where $A_c$ is the collisional age \citep{Maruca2013a}, which estimates the number of collisional timescales for protons. To account for heliospheric sector structure in the magnetic field measurements, we first calculate the Parker-spiral angle, $\theta_{rB}=\arctan{(B_{0,T}/B_{0,R})}$, where $B_{0,R}$ and $B_{0,T}$ are the average components of $\mathbf{B}_0$ over $92\;\textrm{s}$ periods. If $\left<\theta_{rB}\right>$ over a two day period exceeds $45^\circ$ from the radial direction, we reverse the signs of the $B_{0,R}$ and $B_{0,T}$ components so that inwards fields are rotated outwards. This procedure removes the inversion of the sign of magnetic helicity due to the direction $\mathbf{B}_0$ with respect to the Sun.

We transform the 11 Hz magnetic field data into field-aligned coordinates (Equation \ref{eqncoords}) using $\mathbf{B}_0$ averaged over 92 s. We compute the continuous wavelet transform \citep{Torrence1998} using a Morlet wavelet to obtain $\mathsf{P}(f)$ as a function of the spacecraft-frame frequency, $f=k_r \left|\mathbf{v}_{sw}\right|/2\pi$. We then calculate magnetic helicity spectra, $\sigma_{xy}$ and $\sigma_{yz}$, using Equation \ref{eqnhelFA}. We average the spectra over 92 s so that a single spectrum overlaps with exactly one SWE measurement, giving a total of 1,696,270 observations, excluding data gaps. This averaging ensures that fluctuations persist for at least several proton gyro-periods, $2\pi/\Omega_p$, to give a clear coherent helicity signature at proton-kinetic scales. Following \citet{Woodham2018}, we estimate the amplitude of $\sigma_{xy}$ and $\sigma_{yz}$ at proton-kinetic scales by fitting a Gaussian to the coherent peak in each spectrum at frequencies $f\sim0.8~\mathrm{Hz}$, close to the Taylor-shifted frequencies for $d_p$ and $\rho_p$. We neglect any peak at $f>f_{noise}$, the frequency at which instrumental noise of the MFI magnetometer becomes significant \footnote{See Appendix in \citet{Woodham2018}.}. We also reject a spectrum if the angular deviation in $\mathbf{B}$ exceeds 15$^\circ$ during the measurement period to ensure that fluctuations at proton-kinetic scales retain their anisotropy with respect to $\mathbf{B}_0$ over 92 s. We designate the amplitude of the peak in each $\sigma_{xy}$ and $\sigma_{yz}$ spectrum as $\sigma_\parallel$ and $\sigma_\perp$, respectively.

%\footnote{We include the probability density distribution of the data for reference in supplementary information, which is in agreement with many other studies \cite{Hellinger2006,Bale2009,Matteini2007}.}

We bin $\sigma_\parallel$ and $\sigma_\perp$ in $\beta_{p,\parallel}$-$T_{p,\perp}/T_{p,\parallel}$ space using logarithmic bins, where $\beta_{p,\parallel}=n_pk_BT_{p,\parallel}/(B_0^2/2\mu_0)$, $n_p$ is the proton density and $B_0=\left|\mathbf{B}_0\right|$. We use equal bin widths of $\Delta \log_{10}(\beta_{p,\parallel}) = \Delta \log_{10}(T_{p,\perp}/T_{p,\parallel}) = 0.05$ and restrict our analysis to $0.01\leq\beta_{p,\parallel}\leq10$ and $0.1\leq T_{p,\perp}/T_{p,\parallel}\leq10$. In our plots, we neglect any bins with fewer than 10 data points to increase the likelihood of statistical convergence. In this parameter space we overplot contours of constant maximum growth rate, $\gamma/\Omega_p$, for the four kinetic instabilities driven by proton temperature anisotropy. We calculate these contours using linear Vlasov-Maxwell theory (see \citet{Maruca2012} and references therein).

\section{\label{sec:level4}Results \& Discussion}

The presence of an $\alpha$-particle drift can break the symmetry of the proton VDFs, leading to a preferential driving of waves generated by anisotropy-driven AIC and parallel firehose instabilities. Linear Vlasov-Maxwell theory shows that the growth rates of AIC and FMW modes are greater in the anti-sunward and sunward directions, respectively, for $\mathbf{v}_d$ directed anti-sunward \citep{Podesta2011a,Podesta2011}. The propagation of AIC and FMW modes in different directions therefore leads to sign changes in the helicity of these waves when $\sigma_{\parallel}$ is transformed from the plasma-frame to the spacecraft-frame. We summarise the possible cases for the sign of $\sigma_\parallel$ in Table \ref{tab:table1}. For example, if $\mathbf{B}_0$ is directed anti-sunward, then left-handed AIC modes will have $\sigma_\parallel<0$ or $\sigma_\parallel>0$ if they propagate anti-sunward or sunward, respectively. By accounting for sector structure (see Section \ref{sec:Method}), our resulting dataset is consistent with either case I or II from Table \ref{tab:table1}, removing ambiguity in the sign of $\sigma_\parallel$ due to the direction of $\mathbf{B}_0$. Therefore, we hypothesise that $\sigma_\parallel<0$ for both AIC and FMW modes present at proton-kinetic scales in the solar wind.

\begin{figure}[!t]
	\centering
	\includegraphics[width=0.5\textwidth]{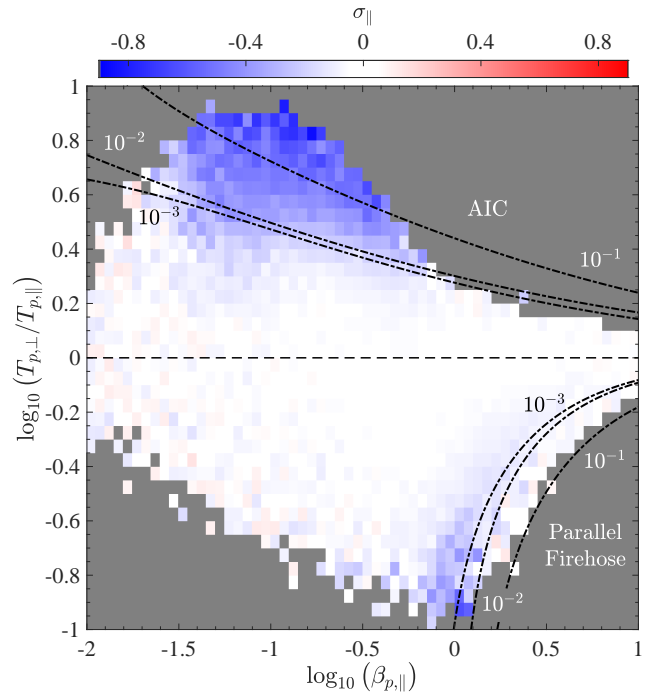} 
	\caption{Median value of $\sigma_\parallel$ across $\beta_{p,\parallel}$-$T_{p,\perp}/T_{p,\parallel}$ space. We overplot contours of different constant maximum growth rates, $\gamma/\Omega_p$, for the AIC and parallel firehose instabilities.} %We include the originial plot without accounting for the Doppler shift in supplementary materials.
	\label{fig:1}
\end{figure}

\begin{table}[b]
	\caption{\label{tab:table1} The four cases for $\mathbf{k}\cdot\mathbf{B}_0$ in the solar wind due to sector structure.}
	\begin{ruledtabular}
		\begin{tabular}{ccccc}
			\textrm{}&
			\textrm{I}&
			\textrm{II}&
			\textrm{III}&
			\textrm{IV}\\
			\colrule
			$\mathbf{B}_0$ & Out & Out & In & In\\
			$\mathbf{k}$ & Out & In & Out & In\\
			\colrule
			$\sigma_{L}$\footnote{Here, $\sigma_L$ and $\sigma_{R}$ give the sign of the magnetic helicity due to left-handed and right-handed fluctuations, respectively. The +(-) sign designates a positive (negative) helicity.} & $-$ & $+$ & $+$ & $-$\\
			$\sigma_{R}$ & $+$ & $-$ & $-$ & $+$\\
		\end{tabular}
	\end{ruledtabular}
\end{table}

To test this hypothesis, we plot in Figure \ref{fig:1} the median $\sigma_\parallel$-value across the $\beta_{p,\parallel}$-$T_{p,\perp}/T_{p,\parallel}$ plane. The black dashed-lines show contours of constant $\gamma/\Omega_p$ for the AIC and parallel firehose instabilities, which have greater growth rates along $\mathbf{B}_0$. We see that the solar wind plasma occupies a significant extent of parameter space in the regions unstable to both the AIC and parallel firehose instabilities, as widely reported in the literature \citep[e.g.,][]{Hellinger2006,Bale2009,Maruca2012}. In these regions of parameter space, we see two distinct signatures at $T_{p,\perp}>T_{p,\parallel}$ and $T_{p,\perp}<T_{p,\parallel}$ where the median value of $\sigma_\parallel$ assumes more negative values. These signatures indicate the presence of coherent fluctuations that we attribute to growing modes from these instabilities. The minimum helicity is about $\sigma_\parallel\simeq-0.6$ for the AIC modes and $\sigma_\parallel\simeq-0.4$ for the FMW modes. Since $\sigma_{\parallel}<0$ corresponds to left-handed helicity in the spacecraft frame, Figure \ref{fig:1} indicates that AIC modes are preferentially driven anti-sunward, and that FMW modes are preferentially driven sunward. We confirm that these fluctuations have median $k_\|d_p\sim1$ at the peak value of $\sigma_\|$ from Figure \ref{fig:1} (not shown here), in agreement with the predictions for linear growth of AIC and FMW modes \citep[e.g., see][]{Klein2015}. This result is consistent with our predictions as well as observations of quasi-parallel propagating waves in the solar wind \citep{Tsurutani1994,Jian2009,Jian2010,He2011,He2012a,He2012,Podesta2011,Klein2014,Jian2014,Bruno2015,He2015,Telloni2015,Telloni2016,Wicks2016,Zhao2018,Zhao2019}. Away from the unstable regions of the parallel instabilities in parameter space and close to $T_{p,\perp}\simeq T_{p,\parallel}$, $\sigma_\parallel\simeq0$, which indicates a lack of coherence in $\mathbf{B}$.

\begin{figure}%[!b]
	\centering
	\includegraphics[width=0.5\textwidth]{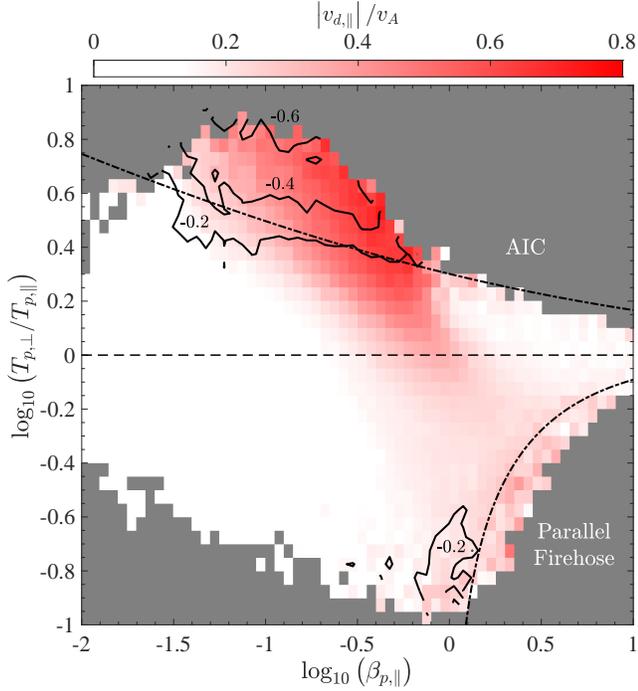} 
	\caption{Median parallel $\alpha$-proton drift, $\left|v_{d,\parallel}\right|/v_A$, across $\beta_{p,\parallel}$-$T_{p,\perp}/T_{p,\parallel}$ space. We overplot contours of constant maximum growth rate, $\gamma/\Omega_p=10^{-2}$, for the AIC and parallel firehose instabilities. We also show contours of constant $\sigma_{\parallel}$ from Figure \ref{fig:1} for reference.}
	\label{fig:3}
\end{figure}

In Figure \ref{fig:3}, we plot the median value of $\left|v_{d,\parallel}\right|/v_A$, the $\alpha$-particle parallel drift speed normalised by the Alfv\'en speed, across the $\beta_{p,\parallel}$-$T_{p,\perp}/T_{p,\parallel}$ plane. We define $v_{d,\parallel}=\mathbf{v}_d\cdot\mathbf{B}_0/\left|\mathbf{B}_0\right|$. We include contours of constant $\sigma_\parallel$ from Figure \ref{fig:1} to show the correlation between $\left|v_{d,\parallel}\right|/v_A$ and $\sigma_\parallel$ in this space. When a significant drift exists close to the unstable regions of the AIC and parallel firehose instabilities, a coherent signature in $\sigma_\parallel$ also exists. The drift is stronger for $T_{p,\perp}/T_{p,\parallel}>1$, reaching a maximum of $\left|v_{d,\parallel}\right|\simeq0.6\,v_A$ at $\beta_{p,\parallel}>0.1$. This peak in $\left|v_{d,\parallel}\right|$ occurs in the region of parameter space dominated by fast wind streams \citep{Matteini2007}. For parallel firehose unstable regions of the parameter space, the drift is significantly weaker, reaching a maximum of $\left|v_{d,\parallel}\right|\simeq0.2\,v_A$. Therefore, the presence of a drift between ion species in the solar wind can explain the preferential driving associated with the AIC and FMW modes, which is consistent with previous studies \citep{Podesta2011a,Verscharen2013,Klein2018}.

%\footnote{see Supplementary information for a plot of median $v_{sw}$ across $\beta_{p,\parallel}$-$T_{p,\perp}/T_{p,\parallel}$ space.}

\begin{figure}[!b]
	\centering
	\includegraphics[width=0.5\textwidth]{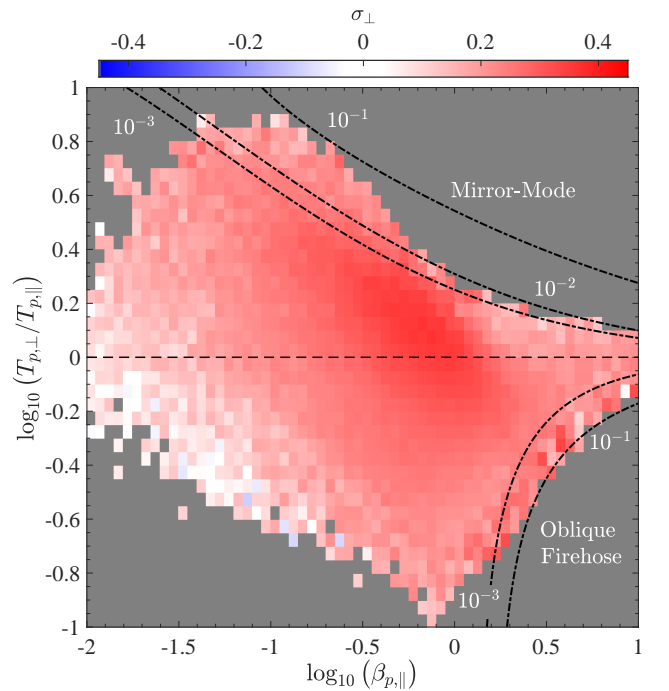} 
	\caption{Median value of $\sigma_\perp$ across $\beta_{p,\parallel}$-$T_{p,\perp}/T_{p,\parallel}$ space. We overplot contours of different constant maximum growth rates, $\gamma/\Omega_p$, for the mirror-mode and oblique firehose instabilities.}
	\label{fig:4}
\end{figure}

Finally, in Figure \ref{fig:4} we plot the median $\sigma_\perp$-value in the same parameter space. We include contours of constant $\gamma/\Omega_p$ for the mirror-mode and oblique firehose instabilities since these have higher growth rates at angles oblique to $\mathbf{B}_0$. Throughout Figure \ref{fig:4}, $\sigma_{\perp}>0$ and peaks at $\sigma_\perp\simeq0.3$, close to $\beta_{p,\parallel}\simeq0.8$ and $T_{p,\perp}\sim T_{p,\parallel}$. This peak lies in a region of parameter space dominated by fast wind, which is typically more Alfv\'enic \citep{Stansby2018}. There is also a small enhancement in the helicity in the unstable region of the oblique firehose instability, suggesting the presence of driven modes with a right-handed helicity in the spacecraft-frame. We do not expect to observe a signature from mirror-modes because they represent structures with $\mathbf{B}$ directed along $\mathbf{B}_0$, which will not be measurable using magnetic helicity.

The lack of a strong dependence of the distribution of $\sigma_\perp$ on $\beta_{p,\parallel}$ and $T_{p,\perp}/T_{p,\parallel}$ implies that the dominant source of these fluctuations is unlikely to be related to kinetic instabilities. Instead, due to the anisotropic nature of the turbulent cascade at proton-kinetic scales, we expect turbulent fluctuations to contribute to $\sigma_\perp$ due to the mode conversion of Alfv\'enic to KAW-like fluctuations at these scales \citep{Markovskii2015}. From linear Vlasov-Maxwell theory, right-handed KAWs with $k_\perp\gg k_\parallel$ at kinetic scales ($k_\perp\rho_p\gtrsim1$) have $\sigma_\perp\simeq1$ for $\mathbf{k}\cdot\mathbf{B}_0>0$ and $\sigma_\perp\simeq-1$ for $\mathbf{k}\cdot\mathbf{B}_0<0$ \citep{Gary1986,Howes2010}. We calculate the median $k_\perp \rho_p$ at the peak value of $\sigma_\perp$ from Figure \ref{fig:4} (not shown here) to assess the scale at which these fluctuations exist, finding that $k_\perp \rho_p\gtrsim1$ where $\sigma_\perp$ is largest in this space. Therefore, Figure \ref{fig:4} is consistent with the presence of outward propagating right-handed fluctuations (Case I from Table \ref{tab:table1}) that we interpret as KAW-like fluctuations from the turbulent cascade. The peak $|\sigma_\perp|<1$ is consistent with the non-linear nature of these fluctuations.

\section{\label{sec:level5}Conclusions}

We use a novel analysis technique to recover information about the wave-vector of solar wind fluctuations using single-point spacecraft measurements. We separate the contributions to magnetic helicity into two components with respect to $\mathbf{B}_0$: one for fluctuations propagating at quasi-parallel angles and the other for those propagating at oblique angles. We analyse over 1.6 million magnetic field and ion spectra from the \textit{Wind} MFI and SWE instruments and quantify the amplitude of the helicity contributions $\sigma_\parallel$ and $\sigma_\perp$ to explore the sources of fluctuations at proton-kinetic scales.

By plotting the median $\sigma_\parallel$-value across $\beta_{p,\parallel}$-$T_{p,\perp}/T_{p,\parallel}$ space, we show that there is a significant negative enhancement in $\sigma_\parallel$ in unstable regions of both the AIC and parallel firehose instabilities. The median value of $\sigma_\parallel$ reaches a minimum of $\sigma_\parallel\simeq-0.6$ at $T_{p,\perp}/T_{p,\parallel}>1$. In the spacecraft-frame, these quasi-parallel propagating fluctuations are left-handed, consistent with left-handed AIC waves propagating anti-sunward for $T_{p,\perp}/T_{p,\parallel}>1$ and right-handed FMW waves propagating sunward in the plasma-frame for $T_{p,\perp}/T_{p,\parallel}<1$. In regions of a negative enhancement in $\sigma_\parallel$, particularly for $T_{p,\perp}/T_{p,\parallel}>1$, we also observe a substantial $\alpha$-particle drift with respect to the proton flow, consistent with predictions \citep{Podesta2011a,Podesta2011}. Elsewhere in $\beta_{p,\parallel}-T_{p,\perp}/T_{p,\parallel}$ space, $\sigma_{\parallel}\simeq0$, which indicates no coherence in $\mathbf{B}$. This result suggests that fluctuations propagating quasi-parallel to $\mathbf{B}_0$ predominantly arise from ion instabilities, consistent with the background solar wind turbulence producing Alfv\'enic fluctuations with $k_\perp\gg k_\parallel$. These results show that instabilities are active and modes generated by them are common in the solar wind.

In addition, we show for the first time that $\sigma_\perp$ is distributed throughout the entire parameter space occupied by the solar wind and peaks at $\sigma_\perp\simeq0.3$. This peak occurs at $T_{p,\perp}\simeq T_{p,\parallel}$ and $\beta_{p,\parallel}\simeq0.8$, which is strongest in a region of $\beta_{p,\parallel}$-$T_{p,\perp}/T_{p,\parallel}$ space dominated by fast wind, suggesting that these fluctuations are more Alfv\'enic. Since $\sigma_\perp>0$ and shows little dependence on $\beta_{p,\parallel}$ and $T_{p,\perp}/T_{p,\parallel}$, this signature is consistent with anisotropic fluctuations from the turbulent cascade with significant $k_\perp$ at proton-kinetic scales. While we interpret these fluctuations as KAW-like modes, we do not rule out that other non-linear turbulent fluctuations or structures contribute to this helicity signal. We conjecture that these fluctuations are insensitive to proton temperature anisotropy and instability growth, in agreement with \citet{Klein2015}. Furthermore, since the unstable AIC and FMW modes do not appear to interact with the turbulent cascade, and there is no evidence of helicity from turbulent fluctuations with significant $k_\parallel$, we provide evidence for a very limited role of quasi-parallel propagating fluctuations in solar wind turbulence at proton-kinetic scales.

Our results provide evidence that the behaviour of fluctuations at proton-kinetic scales is independent of the origin and macroscopic properties of the solar wind. For example, left-handed AIC modes are generated in both fast and slow wind streams, depending only on the local properties of the plasma such as proton temperature anisotropy and the presence of $\alpha$-particle differential flow. In addition, we find no evidence of a parallel-propagating contribution to the helicity from the turbulence cascade at these scales in the stable parameter regime. Any Alfv\'enic fluctuations from the cascade with a significant $k_\|$ would produce a signature in Figure \ref{fig:1} with a similar distribution to the right-handed signal in Figure \ref{fig:4}. Therefore, we can rule out the existence of imbalanced fluctuations with $k_{\parallel} \gtrsim k_{\perp}$ that are not created by instabilities. This result constrains theories of turbulence in the solar wind and their implications for energy transport and dissipation.

The method we employ here can be applied \textit{Parker Solar Probe} and \textit{Solar Orbiter} data to explore the role of fluctuations at kinetic scales in the corona and their evolution with increasing heliocentric distance. This will help us to diagnose the source and nature of the fluctuations that are crucial for the acceleration and heating of the solar wind.

\begin{acknowledgments}
	We thank Kris Klein for useful scientific discussions and Mike Stevens for assistance with the SWE Faraday cup data. Data from the \textit{Wind} spacecraft are obtained from the \href{http://spdf.gsfc.nasa.gov}{SPDF web-site}. LDW is funded by an STFC Studentship; DV is supported by STFC Ernest Rutherford Fellowship ST/P003826/1; RTW and CJO are supported by the STFC consolidated grants to UCL/MSSL, ST/N000722/1 and ST/S000240/1. BLA is supported by NASA grant 80NSSC18K0986.
	%ORCID: \href{https://orcid.org/0000-0003-2845-4250}{Lloyd Woodham},  \href{https://orcid.org/0000-0002-0622-5302}{Robert Wicks}, \href{https://orcid.org/0000-0002-0497-1096}{Daniel Verscharen}, \href{https://orcid.org/0000-0002-5982-4667}{Christopher Owen}, \href{https://orcid.org/0000-0002-2229-5618}{Bennett Maruca}, \href{https://orcid.org/0000-0001-6673-3432}{Benjamin Alterman}
\end{acknowledgments}

\bibliographystyle{yahapj}
\bibliography{Bibliography}

\begin{thebibliography}{}
\providecommand\natexlab[1]{#1}
\providecommand\JournalTitle[1]{#1}

\bibitem[{Acu{\~{n}}a {et~al.}(1995)Acu{\~{n}}a, Ogilvie, Baker, Curtis,
  Fairfield, \& Mish}]{Acuna1995}
Acu{\~{n}}a, M.~H., Ogilvie, K.~W., Baker, D.~N., {et~al.} 1995,
  \href{http://dx.doi.org/10.1007/BF00751323}{\JournalTitle{Space Science
  Reviews}, 71, 5}

\bibitem[{Alexandrova {et~al.}(2013)Alexandrova, Chen, Sorriso-Valvo, Horbury,
  \& Bale}]{Alexandrova2013}
Alexandrova, O., Chen, C. H.~K., Sorriso-Valvo, L., Horbury, T.~S., \& Bale,
  S.~D. 2013,
  \href{http://dx.doi.org/10.1007/s11214-013-0004-8}{\JournalTitle{Space
  Science Reviews}, 178, 101}

\bibitem[{Alterman {et~al.}(2018)Alterman, Kasper, Stevens, \&
  Koval}]{Alterman2018}
Alterman, B.~L., Kasper, J.~C., Stevens, M.~L., \& Koval, A. 2018,
  \href{http://dx.doi.org/10.3847/1538-4357/aad23f}{\JournalTitle{The
  Astrophysical Journal}, 864, 112}

\bibitem[{Bale {et~al.}(2009)Bale, Kasper, Howes, Quataert, Salem, \&
  Sundkvist}]{Bale2009}
Bale, S.~D., Kasper, J.~C., Howes, G.~G., {et~al.} 2009,
  \href{http://dx.doi.org/10.1103/PhysRevLett.103.211101}{\JournalTitle{Physical
  Review Letters}, 103, 211101}

\bibitem[{Bale {et~al.}(2005)Bale, Kellogg, Mozer, Horbury, \& Reme}]{Bale2005}
Bale, S.~D., Kellogg, P.~J., Mozer, F.~S., Horbury, T.~S., \& Reme, H. 2005,
  \href{http://dx.doi.org/10.1103/PhysRevLett.94.215002}{\JournalTitle{Physical
  Review Letters}, 94, 215002}

\bibitem[{Batchelor(1970)}]{Batchelor1970}
Batchelor, G.~K. 1970, {The Theory of Homogenous Turbulence} (Cambridge
  University Press)

\bibitem[{Bourouaine {et~al.}(2013)Bourouaine, Verscharen, Chandran, Maruca, \&
  Kasper}]{Bourouaine2013}
Bourouaine, S., Verscharen, D., Chandran, B.~D., Maruca, B.~A., \& Kasper,
  J.~C. 2013,
  \href{http://dx.doi.org/10.1088/2041-8205/777/1/L3}{\JournalTitle{Astrophysical
  Journal Letters}, 777, L3}

\bibitem[{Bruno \& Carbone(2013)}]{Bruno2013}
Bruno, R., \& Carbone, V. 2013,
  \href{http://dx.doi.org/10.12942/lrsp-2013-2}{\JournalTitle{Living Reviews in
  Solar Physics}, 10, 2}

\bibitem[{Bruno \& Telloni(2015)}]{Bruno2015}
Bruno, R., \& Telloni, D. 2015,
  \href{http://dx.doi.org/10.1088/2041-8205/811/2/L17}{\JournalTitle{The
  Astrophysical Journal Letters}, 811, L17}

\bibitem[{Chen(2016)}]{Chen2016a}
Chen, C. H.~K. 2016,
  \href{http://dx.doi.org/10.1017/S0022377816001124}{\JournalTitle{Journal of
  Plasma Physics}, 82, 535820602}

\bibitem[{Chen {et~al.}(2010)Chen, Horbury, Schekochihin, Wicks, Alexandrova,
  \& Mitchell}]{Chen2010a}
Chen, C. H.~K., Horbury, T.~S., Schekochihin, A.~A., {et~al.} 2010,
  \href{http://dx.doi.org/10.1103/PhysRevLett.104.255002}{\JournalTitle{Physical
  Review Letters}, 104, 255002}

\bibitem[{Chen {et~al.}(2019)Chen, Klein, \& Howes}]{Chen2018}
Chen, C. H.~K., Klein, K.~G., \& Howes, G.~G. 2019,
  \href{http://dx.doi.org/10.1038/s41467-019-08435-3}{\JournalTitle{Nature
  Communications}, 10, 740}

\bibitem[{Chen {et~al.}(2012)Chen, Mallet, Schekochihin, Horbury, Wicks, \&
  Bale}]{Chen2012a}
Chen, C. H.~K., Mallet, A., Schekochihin, A.~A., {et~al.} 2012,
  \href{http://dx.doi.org/10.1088/0004-637X/758/2/120}{\JournalTitle{The
  Astrophysical Journal}, 758, 120}

\bibitem[{Chen {et~al.}(2011)Chen, Mallet, Yousef, Schekochihin, \&
  Horbury}]{Chen2011}
Chen, C. H.~K., Mallet, A., Yousef, T.~A., Schekochihin, A.~A., \& Horbury,
  T.~S. 2011,
  \href{http://dx.doi.org/10.1111/j.1365-2966.2011.18933.x}{\JournalTitle{Monthly
  Notices of the Royal Astronomical Society}, 415, 3219}

\bibitem[{Chew {et~al.}(1956)Chew, Goldberger, \& Low}]{Chew1956}
Chew, G.~F., Goldberger, M.~L., \& Low, F.~E. 1956,
  \href{http://dx.doi.org/10.1098/rspa.1956.0116}{\JournalTitle{Proceedings of
  the Royal Society A: Mathematical and Physical Sciences}, 236}

\bibitem[{Gary(1986)}]{Gary1986}
Gary, S.~P. 1986,
  \href{http://dx.doi.org/10.1017/S0022377800011442}{\JournalTitle{Journal of
  Plasma Physics}, 35, 431}

\bibitem[{Gary {et~al.}(2015)Gary, Jian, Broiles, Stevens, Podesta, \&
  Kasper}]{Gary2015a}
Gary, S.~P., Jian, L.~K., Broiles, T.~W., {et~al.} 2015,
  \href{http://dx.doi.org/10.1002/2015JA021935}{\JournalTitle{Journal of
  Geophysical Research: Space Physics}, 121, 30}

\bibitem[{He {et~al.}(2011)He, Marsch, Tu, Yao, \& Tian}]{He2011}
He, J., Marsch, E., Tu, C.~Y., Yao, S., \& Tian, H. 2011,
  \href{http://dx.doi.org/10.1088/0004-637X/731/2/85}{\JournalTitle{The
  Astrophysical Journal}, 731, 85}

\bibitem[{He {et~al.}(2012{\natexlab{a}})He, Tu, Marsch, \& Yao}]{He2012a}
He, J., Tu, C.~Y., Marsch, E., \& Yao, S. 2012{\natexlab{a}},
  \href{http://dx.doi.org/10.1088/2041-8205/745/1/L8}{\JournalTitle{The
  Astrophysical Journal Letters}, 745, L8}

\bibitem[{He {et~al.}(2012{\natexlab{b}})He, Tu, Marsch, \& Yao}]{He2012}
---. 2012{\natexlab{b}},
  \href{http://dx.doi.org/10.1088/0004-637X/749/1/86}{\JournalTitle{The
  Astrophysical Journal}, 749, 86}

\bibitem[{He {et~al.}(2015)He, Wang, Tu, Marsch, \& Zong}]{He2015}
He, J., Wang, L., Tu, C.~Y., Marsch, E., \& Zong, Q. 2015,
  \href{http://dx.doi.org/10.1088/2041-8205/800/2/L31}{\JournalTitle{The
  Astrophysical Journal Letters}, 800, L31}

\bibitem[{Hellinger {et~al.}(2006)Hellinger, Tr{\'{a}}vn{\'{i}}{\v{c}}ek,
  Kasper, \& Lazarus}]{Hellinger2006}
Hellinger, P., Tr{\'{a}}vn{\'{i}}{\v{c}}ek, P.~M., Kasper, J.~C., \& Lazarus,
  A.~J. 2006,
  \href{http://dx.doi.org/10.1029/2006GL025925}{\JournalTitle{Geophysical
  Research Letters}, 33, L09101}

\bibitem[{Horbury {et~al.}(2008)Horbury, Forman, \& Oughton}]{Horbury2008}
Horbury, T.~S., Forman, M.~A., \& Oughton, S. 2008,
  \href{http://dx.doi.org/10.1103/PhysRevLett.101.175005}{\JournalTitle{Physical
  Review Letters}, 101, 175005}

\bibitem[{Horbury {et~al.}(2012)Horbury, Wicks, \& Chen}]{Horbury2012}
Horbury, T.~S., Wicks, R.~T., \& Chen, C. H.~K. 2012,
  \href{http://dx.doi.org/10.1007/s11214-011-9821-9}{\JournalTitle{Space
  Science Reviews}, 172, 325}

\bibitem[{Howes {et~al.}(2008)Howes, Dorland, Cowley, Hammett, Quataert,
  Schekochihin, \& Tatsuno}]{Howes2008b}
Howes, G.~G., Dorland, W., Cowley, S.~C., {et~al.} 2008,
  \href{http://dx.doi.org/10.1103/PhysRevLett.100.065004}{\JournalTitle{Physical
  Review Letters}, 100, 065004}

\bibitem[{Howes \& Quataert(2010)}]{Howes2010}
Howes, G.~G., \& Quataert, E. 2010,
  \href{http://dx.doi.org/10.1088/2041-8205/709/1/L49}{\JournalTitle{The
  Astrophysical Journal Letters}, 709, L49}

\bibitem[{Jian {et~al.}(2010)Jian, Russell, Luhmann, Anderson, Boardsen,
  Strangeway, Cowee, \& Wennmacher}]{Jian2010}
Jian, L.~K., Russell, C.~T., Luhmann, J.~G., {et~al.} 2010,
  \href{http://dx.doi.org/10.1029/2010JA015737}{\JournalTitle{Journal of
  Geophysical Research: Space Physics}, 115, A12115}

\bibitem[{Jian {et~al.}(2009)Jian, Russell, Luhmann, Strangeway, Leisner, \&
  Galvin}]{Jian2009}
---. 2009,
  \href{http://dx.doi.org/10.1088/0004-637X/701/2/L105}{\JournalTitle{The
  Astrophysical Journal}, 701, L105}

\bibitem[{Jian {et~al.}(2014)Jian, Wei, Russell, Luhmann, Klecker, Omidi,
  Isenberg, Goldstein, Figueroa-Vi{\~{n}}as, \& Blanco-Cano}]{Jian2014}
Jian, L.~K., Wei, H.~Y., Russell, C.~T., {et~al.} 2014,
  \href{http://dx.doi.org/10.1088/0004-637X/786/2/123}{\JournalTitle{The
  Astrophysical Journal}, 786, 123}

\bibitem[{Kasper {et~al.}(2002)Kasper, Lazarus, \& Gary}]{Kasper2002a}
Kasper, J.~C., Lazarus, A.~J., \& Gary, S.~P. 2002,
  \href{http://dx.doi.org/10.1029/2002GL015128}{\JournalTitle{Geophysical
  Research Letters}, 29, 1839}

\bibitem[{Kasper {et~al.}(2008)Kasper, Lazarus, \& Gary}]{Kasper2008}
---. 2008,
  \href{http://dx.doi.org/10.1103/PhysRevLett.101.261103}{\JournalTitle{Physical
  Review Letters}, 101, 261103}

\bibitem[{Kasper {et~al.}(2006)Kasper, Lazarus, Steinberg, Ogilvie, \&
  Szabo}]{Kasper2006}
Kasper, J.~C., Lazarus, A.~J., Steinberg, J.~T., Ogilvie, K.~W., \& Szabo, A.
  2006, \href{http://dx.doi.org/10.1029/2005JA011442}{\JournalTitle{Journal of
  Geophysical Research: Space Physics}, 111, A03105}

\bibitem[{Kasper {et~al.}(2013)Kasper, Maruca, Stevens, \&
  Zaslavsky}]{Kasper2013}
Kasper, J.~C., Maruca, B.~A., Stevens, M.~L., \& Zaslavsky, A. 2013,
  \href{http://dx.doi.org/10.1103/PhysRevLett.110.091102}{\JournalTitle{Physical
  Review Letters}, 110, 091102}

\bibitem[{Kasper {et~al.}(2017)Kasper, Klein, Weber, Maksimovic, Zaslavsky,
  Bale, Maruca, Stevens, \& Case}]{Kasper2017}
Kasper, J.~C., Klein, K.~G., Weber, T., {et~al.} 2017,
  \href{http://dx.doi.org/10.3847/1538-4357/aa84b1}{\JournalTitle{The
  Astrophysical Journal}, 849, 126}

\bibitem[{Klein {et~al.}(2018)Klein, Alterman, Stevens, Vech, \&
  Kasper}]{Klein2018}
Klein, K.~G., Alterman, B.~L., Stevens, M.~L., Vech, D., \& Kasper, J.~C. 2018,
  \href{http://dx.doi.org/10.1103/PhysRevLett.120.205102}{\JournalTitle{Physical
  Review Letters}, 120, 205102}

\bibitem[{Klein \& Howes(2015)}]{Klein2015}
Klein, K.~G., \& Howes, G.~G. 2015,
  \href{http://dx.doi.org/10.1063/1.4914933}{\JournalTitle{Physics of Plasmas},
  22, 032903}

\bibitem[{Klein {et~al.}(2014)Klein, Howes, TenBarge, \& Podesta}]{Klein2014}
Klein, K.~G., Howes, G.~G., TenBarge, J.~M., \& Podesta, J.~J. 2014,
  \href{http://dx.doi.org/10.1088/0004-637X/785/2/138}{\JournalTitle{The
  Astrophysical Journal}, 785, 138}

\bibitem[{Koval \& Szabo(2013)}]{Koval2013}
Koval, A., \& Szabo, A. 2013, \href{http://dx.doi.org/10.1063/1.4811025}{in AIP
  Conference Proceedings, Vol. 1539}, 211

\bibitem[{Lacombe {et~al.}(2017)Lacombe, Alexandrova, \&
  Matteini}]{Lacombe2017}
Lacombe, C., Alexandrova, O., \& Matteini, L. 2017,
  \href{http://dx.doi.org/10.3847/1538-4357/aa8c06}{\JournalTitle{The
  Astrophysical Journal}, 848, 45}

\bibitem[{Leamon {et~al.}(1999)Leamon, Smith, Ness, \& Wong}]{Leamon1999}
Leamon, R.~J., Smith, C.~W., Ness, N.~F., \& Wong, H.~K. 1999,
  \href{http://dx.doi.org/10.1029/1999JA900158}{\JournalTitle{Journal of
  Geophysical Research: Space Physics}, 104, 22331}

\bibitem[{Lepping {et~al.}(1995)Lepping, Acu{\~{n}}a, Burlaga, Farrell, Slavin,
  Schatten, Mariani, Ness, Neubauer, Whang, Byrnes, Kennon, Panetta, Scheifele,
  \& Worley}]{Lepping1995}
Lepping, R.~P., Acu{\~{n}}a, M.~H., Burlaga, L.~F., {et~al.} 1995,
  \href{http://dx.doi.org/10.1007/BF00751330}{\JournalTitle{Space Science
  Reviews}, 71, 207}

\bibitem[{Markovskii {et~al.}(2015)Markovskii, Vasquez, \&
  Smith}]{Markovskii2015}
Markovskii, S.~A., Vasquez, B.~J., \& Smith, C.~W. 2015,
  \href{http://dx.doi.org/10.1088/0004-637X/806/1/78}{\JournalTitle{The
  Astrophysical Journal}, 806, 78}

\bibitem[{Marsch(2012)}]{Marsch2012}
Marsch, E. 2012,
  \href{http://dx.doi.org/10.1007/s11214-010-9734-z}{\JournalTitle{Space
  Science Reviews}, 172, 23}

\bibitem[{Maruca {et~al.}(2013)Maruca, Bale, Sorriso-Valvo, Kasper, \&
  Stevens}]{Maruca2013a}
Maruca, B.~A., Bale, S.~D., Sorriso-Valvo, L., Kasper, J.~C., \& Stevens, M.~L.
  2013,
  \href{http://dx.doi.org/10.1103/PhysRevLett.111.241101}{\JournalTitle{Physical
  Review Letters}, 111, 241101}

\bibitem[{Maruca {et~al.}(2012)Maruca, Kasper, \& Gary}]{Maruca2012}
Maruca, B.~A., Kasper, J.~C., \& Gary, S.~P. 2012,
  \href{http://dx.doi.org/10.1088/0004-637X/748/2/137}{\JournalTitle{The
  Astrophysical Journal}, 748, 137}

\bibitem[{Matteini {et~al.}(2007)Matteini, Landi, Hellinger, Pantellini,
  Maksimovic, Velli, Goldstein, \& Marsch}]{Matteini2007}
Matteini, L., Landi, S., Hellinger, P., {et~al.} 2007,
  \href{http://dx.doi.org/10.1029/2007GL030920}{\JournalTitle{Geophysical
  Research Letters}, 34, L20105}

\bibitem[{Matthaeus \& Goldstein(1982)}]{Matthaeus1982b}
Matthaeus, W.~H., \& Goldstein, M.~L. 1982,
  \href{http://dx.doi.org/doi:10.1029/JA087iA12p10347}{\JournalTitle{Journal of
  Geophysical Research}, 87, 10347}

\bibitem[{Matthaeus {et~al.}(1982)Matthaeus, Goldstein, \&
  Smith}]{Matthaeus1982a}
Matthaeus, W.~H., Goldstein, M.~L., \& Smith, C.~W. 1982,
  \href{http://dx.doi.org/10.1103/PhysRevLett.48.1256}{\JournalTitle{Physical
  Review Letters}, 48, 1256}

\bibitem[{Montgomery \& Turner(1981)}]{Montgomery1981}
Montgomery, M.~D., \& Turner, L. 1981,
  \href{http://dx.doi.org/10.1063/1.863455}{\JournalTitle{Physics of Fluids},
  24, 825}

\bibitem[{Neugebauer {et~al.}(1994)Neugebauer, Goldstein, Bame, \&
  Feldman}]{Neugebauer1994}
Neugebauer, M., Goldstein, B.~E., Bame, S.~J., \& Feldman, W.~C. 1994,
  \href{http://dx.doi.org/10.1029/93JA02615}{\JournalTitle{Journal of
  Geophysical Research: Space Physics}, 99, 2505}

\bibitem[{Neugebauer {et~al.}(1996)Neugebauer, Goldstein, Smith, \&
  Feldman}]{Neugebauer1996}
Neugebauer, M., Goldstein, B.~E., Smith, E.~J., \& Feldman, W.~C. 1996,
  \href{http://dx.doi.org/10.1029/96JA01406}{\JournalTitle{Journal of
  Geophysical Research}, 101, 17047}

\bibitem[{Ogilvie {et~al.}(1995)Ogilvie, Chornay, Fritzenreiter, Hunsaker,
  Keller, Lobell, Miller, Scudder, Sittler, Torbert, Bodet, Needell, Lazarus,
  Steinberg, Tappan, Mavretic, \& Gergin}]{Ogilvie1995}
Ogilvie, K.~W., Chornay, D.~J., Fritzenreiter, R.~J., {et~al.} 1995,
  \href{http://dx.doi.org/10.1007/BF00751326}{\JournalTitle{Space Science
  Reviews}, 71, 55}

\bibitem[{Podesta \& Gary(2011{\natexlab{a}})}]{Podesta2011a}
Podesta, J.~J., \& Gary, S.~P. 2011{\natexlab{a}},
  \href{http://dx.doi.org/10.1088/0004-637X/742/1/41}{\JournalTitle{The
  Astrophysical Journal}, 742, 41}

\bibitem[{Podesta \& Gary(2011{\natexlab{b}})}]{Podesta2011}
---. 2011{\natexlab{b}},
  \href{http://dx.doi.org/10.1088/0004-637X/734/1/15}{\JournalTitle{The
  Astrophysical Journal}, 734, 15}

\bibitem[{Sahraoui {et~al.}(2010)Sahraoui, Goldstein, Belmont, Canu, \&
  Rezeau}]{Sahraoui2010}
Sahraoui, F., Goldstein, M.~L., Belmont, G., Canu, P., \& Rezeau, L. 2010,
  \href{http://dx.doi.org/10.1103/PhysRevLett.105.131101}{\JournalTitle{Physical
  Review Letters}, 105, 131101}

\bibitem[{Stansby {et~al.}(2019)Stansby, Horbury, \& Matteini}]{Stansby2018}
Stansby, D., Horbury, T.~S., \& Matteini, L. 2019,
  \href{http://dx.doi.org/10.1093/mnras/sty2814}{\JournalTitle{Monthly Notices
  of the Royal Astronomical Society}, 482, 1706}

\bibitem[{Steinberg {et~al.}(1996)Steinberg, Lazarus, Ogilvie, Lepping, \&
  Byrnes}]{Steinberg1996}
Steinberg, J.~T., Lazarus, A.~J., Ogilvie, K.~W., Lepping, R.~P., \& Byrnes,
  J.~B. 1996,
  \href{http://dx.doi.org/10.1029/96GL00628}{\JournalTitle{Geophysical Research
  Letters}, 23, 1183}

\bibitem[{Taylor(1938)}]{Taylor1938}
Taylor, G.~I. 1938,
  \href{http://dx.doi.org/10.1098/rspa.1938.0032}{\JournalTitle{Proceedings of
  the Royal Society A: Mathematical and Physical Sciences}, 164, 476}

\bibitem[{Telloni \& Bruno(2016)}]{Telloni2016}
Telloni, D., \& Bruno, R. 2016,
  \href{http://dx.doi.org/10.1093/mnrasl/slw135}{\JournalTitle{Monthly Notices
  of the Royal Astronomical Society: Letters}, 463, L79}

\bibitem[{Telloni {et~al.}(2015)Telloni, Bruno, \& Trenchi}]{Telloni2015}
Telloni, D., Bruno, R., \& Trenchi, L. 2015,
  \href{http://dx.doi.org/10.1088/0004-637X/805/1/46}{\JournalTitle{The
  Astrophysical Journal}, 805, 46}

\bibitem[{Torrence \& Compo(1998)}]{Torrence1998}
Torrence, C., \& Compo, G.~P. 1998,
  \href{http://dx.doi.org/10.1175/1520-0477(1998)079<0061:APGTWA>2.0.CO;2}{\JournalTitle{Bulletin
  of the American Meteorological Society}, 79, 61}

\bibitem[{Tsurutani {et~al.}(1994)Tsurutani, Arballo, Mok, Smith, Mason, \&
  Tan}]{Tsurutani1994}
Tsurutani, B.~T., Arballo, J.~K., Mok, J., {et~al.} 1994,
  \href{http://dx.doi.org/10.1029/94GL00566}{\JournalTitle{Geophysical Research
  Letters}, 21, 633}

\bibitem[{Verscharen {et~al.}(2013)Verscharen, Bourouaine, \&
  Chandran}]{Verscharen2013}
Verscharen, D., Bourouaine, S., \& Chandran, B.~D. 2013,
  \href{http://dx.doi.org/10.1088/0004-637X/773/2/163}{\JournalTitle{Astrophysical
  Journal}, 773, 163}

\bibitem[{Verscharen {et~al.}(2019)Verscharen, Klein, \&
  Maruca}]{Verscharen2019}
Verscharen, D., Klein, K.~G., \& Maruca, B.~A. 2019,
  \href{http://arxiv.org/abs/1902.03448}{\JournalTitle{Submitted to Living Reviews in Solar
  Physics}},
  \href{http://arxiv.org/abs/1902.03448}{{\sffamily arXiv:1902.03448}}

\bibitem[{Wicks {et~al.}(2012)Wicks, Forman, Horbury, \& Oughton}]{Wicks2012}
Wicks, R.~T., Forman, M.~A., Horbury, T.~S., \& Oughton, S. 2012,
  \href{http://dx.doi.org/10.1088/0004-637X/746/1/103}{\JournalTitle{The
  Astrophysical Journal}, 746, 103}

\bibitem[{Wicks {et~al.}(2010)Wicks, Horbury, Chen, \&
  Schekochihin}]{Wicks2010a}
Wicks, R.~T., Horbury, T.~S., Chen, C. H.~K., \& Schekochihin, A.~A. 2010,
  \href{http://dx.doi.org/10.1111/j.1745-3933.2010.00898.x}{\JournalTitle{Monthly
  Notices of the Royal Astronomical Society}, 407, L31}

\bibitem[{Wicks {et~al.}(2016)Wicks, Alexander, Stevens, {Wilson III}, Moya,
  Vi{\~{n}}as, Jian, Roberts, O'Modhrain, Gilbert, \& Zurbuchen}]{Wicks2016}
Wicks, R.~T., Alexander, R.~L., Stevens, M.~L., {et~al.} 2016,
  \href{http://dx.doi.org/10.3847/0004-637X/819/1/6}{\JournalTitle{The
  Astrophysical Journal}, 819, 6}

\bibitem[{Woodham {et~al.}(2018)Woodham, Wicks, Verscharen, \&
  Owen}]{Woodham2018}
Woodham, L.~D., Wicks, R.~T., Verscharen, D., \& Owen, C.~J. 2018,
  \href{http://dx.doi.org/10.3847/1538-4357/aab03d}{\JournalTitle{The
  Astrophysical Journal}, 856, 49}

\bibitem[{Zhao {et~al.}(2018)Zhao, Feng, Wu, Liu, Zhao, Zhao, \&
  Huang}]{Zhao2018}
Zhao, G.~Q., Feng, H.~Q., Wu, D.~J., {et~al.} 2018,
  \href{http://dx.doi.org/10.1002/2017JA024979}{\JournalTitle{Journal of
  Geophysical Research: Space Physics}, 123, 1715}

\bibitem[{Zhao {et~al.}(2019)Zhao, Feng, Wu, Pi, \& Huang}]{Zhao2019}
Zhao, G.~Q., Feng, H.~Q., Wu, D.~J., Pi, G., \& Huang, J. 2019,
  \href{http://dx.doi.org/10.3847/1538-4357/aaf8b8}{\JournalTitle{The
  Astrophysical Journal}, 871, 175}

\end{thebibliography}

%\newpage
%
%\appendix*
%\begin{widetext}
%\section{Supplementary Information}
%
%\begin{figure*}[!t]
%	\centering
%	\includegraphics[width=0.46\textwidth]{Paper0v2.eps}
%	\qquad
%	\includegraphics[width=0.46\textwidth]{Paper01v2.eps}
%	%\includegraphics[width=0.5\textwidth]{Paper01.eps} 
%	\caption{\textit{Left}: Probability density distribution of data across $\beta_{p,\parallel}$-$T_{p,\perp}/T_{p,\parallel}$ space. We overplot contours of constant maximum growth rate, $\gamma/\Omega_p=10^{-2}$, for the proton temperature anisotropy instabilities: AIC, mirror-mode (M), parallel (PF) and oblique firehose (OF). \textit{Right}: Median solar wind speed, $v_{sw}$ across $\beta_{p,\parallel}$-$T_{p,\perp}/T_{p,\parallel}$ space. We overplot contours of $\gamma/\Omega_p=10^{-2}$ for the AIC and PF instabilities.}
%	\label{fig:AA}
%\end{figure*}
%
%In Figure \ref{fig:AA} we show the probability density distribution of the solar wind data used in our study. We overplot contours of constant maximum growth rate, $\gamma/\Omega_p=10^{-2}$, for each of the four kinetic instabilities under discussion: AIC, mirror-mode, parallel and oblique firehose, labelled separately in the plot. We also plot the median solar wind speed, $v_{sw}$ across $\beta_{p,\parallel}$-$T_{p,\perp}/T_{p,\parallel}$ space. See main text for definitions of parameters.
%
%\end{widetext}
\end{document}